\documentclass[sigplan,10pt,screen,nonacm]{acmart}
\usepackage{xspace}
\usepackage[inline]{enumitem}
\usepackage{pifont}
\usepackage{multirow}
\usepackage{subcaption}
\usepackage{microtype}
\usepackage{circledsteps}

\pgfkeys{/csteps/inner color=white}
\pgfkeys{/csteps/outer color=white}
\pgfkeys{/csteps/fill color=black}
\definecolor{heraldBlue}{rgb}{0.0,0.0,0.8}
\definecolor{heraldRed}{rgb}{0.8,0.0,0.0}
\definecolor{heraldGray}{rgb}{0.2,0.2,0.5}
\definecolor{heraldGreen}{rgb}{0.0,0.4,0.0}
\definecolor{heraldPink}{rgb}{0.8,0.1,0.6}
\definecolor{circledOrange}{RGB}{192,79,21}
\def\NOTES{0}

\if \NOTES 1
    \newcommand{\simon}[1]{\textcolor{blue}{\small  [simon: #1]}}
    \newcommand{\rajath}[1]{\textcolor{heraldGreen}{\small [rajath: #1]}}
    \newcommand{\antoine}[1]{\textcolor{orange}{\small [antoine: #1]}}
\else
    \newcommand{\simon}[1]{}
    \newcommand{\rajath}[1]{}
    \newcommand{\antoine}[1]{}
\fi

\newcommand{\sys}{Presto\xspace}
\newcommand{\libsys}{libPresto\xspace}
\newcommand{\libsyszc}{libPresto-ZC\xspace}

\newcommand{\shortsecref}[1]{\S{}\ref{#1}}

\AtBeginDocument{\providecommand\BibTeX{{\normalfont B\kern-0.5em{\scshape i\kern-0.25em b}\kern-0.8em\TeX}}}

\begin{document}
\title{\sys{}: A Match-Action TCP Stack for the Terabit Era}

\author{Rajath Shashidhara}
\orcid{0000-0002-9676-8959}
\affiliation{\institution{University of Washington}\city{Seattle}\state{WA}\country{USA}}
\email{rajaths@cs.washington.edu}
\author{Antoine Kaufmann}
\orcid{0000-0002-6355-2772}
\affiliation{\institution{MPI-SWS}\city{Saarbrücken}\state{Saarland}\country{Germany}}
\email{antoinek@mpi-sws.org}
\author{Simon Peter}
\orcid{0009-0007-4748-8524}
\affiliation{\institution{University of Washington}\city{Seattle}\state{WA}\country{USA}}
\email{simpeter@cs.washington.edu}

\begin{abstract}
We present \sys{}, the first TCP stack that delivers ASIC-class performance
  and energy efficiency on programmable Reconfigurable Match-Action Table (RMT)
  pipelines, providing flexibility while retaining standard TCP semantics and
  POSIX socket compatibility. The key challenge in designing \sys{} is
  reconciling TCP's complex, dependent state updates with RMT's unidirectional,
  lock-step execution model. To overcome this challenge, \sys{} introduces
  three novel techniques: optimistic concurrency (speculative updates validated
  downstream), pseudo-segment injection (circular dependency resolution without
  stalls), and bump-in-the-wire processing (single-pass segment handling).
  Together, these enable TCP retransmission, reassembly, flow, and congestion
  control, as a pipeline of simple match-action operations.

  Our Intel Tofino~2 prototype demonstrates \sys{}'s scalability to terabit
  speeds, flexibility, and robustness to network dynamics.\ \sys{} matches RDMA
  performance and efficiency for both RPC and streaming workloads (including
  NVMe-oF with SPDK), while maintaining TCP/POSIX compatibility.\ \sys{} saves
  up to 16 host CPU cores versus state-of-the-art kernel-bypass TCP, while
  achieving 5$\times$ lower 99.99p tail latency and 2$\times$ better
  throughput-per-watt for key-value stores. At scale, \sys{} drives nearly
  $1$~Bpps at 20~$\mu$s RPC tail latency. Unlike fixed-function offloads,
  \sys{} supports transport evolution through in-data-path extensions
  (selective ACKs, congestion control variants, application co-design for
  shared logs). Finally, \sys{} generalizes to FPGA SmartNICs, outperforming
  Tonic's monolithic design by $3\times$ under equal timing.
\end{abstract}

\maketitle
\hypersetup{pdfauthor={Rajath Shashidhara,
      Antoine Kaufmann, Simon Peter},
  pdfcreator={Rajath Shashidhara,
      Antoine Kaufmann, Simon Peter}}

\section{Introduction}

In modern datacenters, the ``CPU tax'' of networking has become unsustainable.
Even with state-of-the-art kernel-bypass TCP stacks, applications spend up to
48\% of per-packet CPU cycles in the stack~\cite{flextoe}. Yet, software TCP
stacks remain the default due to their broad compatibility, flexibility, and
robust semantics. As networks scale to terabits and latency targets tighten,
operators face an increasingly untenable trade-off among performance,
efficiency, and flexibility. CPU-based stacks cannot keep pace, yielding poor
bandwidth utilization, high energy consumption, and excessive overhead. At the
other extreme, ASIC transports, such as TCP offload engines (TOEs) and remote
direct memory access (RDMA), are highly efficient, but
inflexible and operationally brittle in the
cloud~\cite{1rma,zeronic,singhvi2025falcon}, and are thus largely confined to
high-performance computing and storage
deployments~\cite{bai2023empowering,roce_ai}.
Programmable network processing units (NPUs)~\cite{flextoe} and
FPGAs~\cite{tonic,beehive} only partially bridge this gap and still face
limitations in scalability, complexity, and energy efficiency.

We break this tension by leveraging the Reconfigurable Match-Action Table (RMT)
architecture~\cite{metamorphosis}, widely deployed in programmable network
switches~\cite{tofino2} and SmartNICs~\cite{pensando_elba}. RMT provides
deterministic, energy-efficient line-rate packet processing through a sequence
of match-action stages, while retaining a programmable execution model that has
enabled a wide range of networking
tasks~\cite{conweave,rmtalloc,rmtfairq,bfc,pegasus,mind,concordia,redplane,tea}.
Realizing TCP in RMT, however, is challenging. TCP requires complex,
interdependent state updates, consistent state for bidirectional flow
coordination, and buffering for
segment reassembly and retransmission, all ill-suited to RMT's strictly
unidirectional pipeline with limited local state and tight timing constraints.
Bridging this mismatch would unlock ASIC performance and efficiency with
software flexibility.

We present \sys{}, the first TCP stack tailored to the RMT architecture.\
\sys{} reimagines TCP as a pipeline of lightweight match-action operations,
enabling line-rate processing without buffering and delivering practical TCP at
terabit speeds while preserving high throughput, low latency, energy
efficiency, flexibility, and interoperability. To reconcile TCP's complex state
dependencies and RMT's unidirectional flow, \sys{} introduces:
\begin{enumerate*}[label={(\arabic*)}]
  \item \emph{optimistic concurrency with deferred validation} to resolve
        \emph{forward read dependencies}, where earlier stages depend on state
        available only in later stages, by performing speculative updates validated downstream;
  \item \emph{pseudo segment injection}, which eliminates \emph{circular write
          dependencies} without side-effects or pipeline stalls; and
  \item a \emph{bump-in-the-wire} design that avoids recirculation and performs
        stateful TCP segment processing in a single pipeline pass.
\end{enumerate*}
Together, these principles enable terabit-scale performance, while maintaining
TCP semantics and POSIX socket compatibility.

\begin{figure*}
  \centering
  \includegraphics[width=\textwidth]{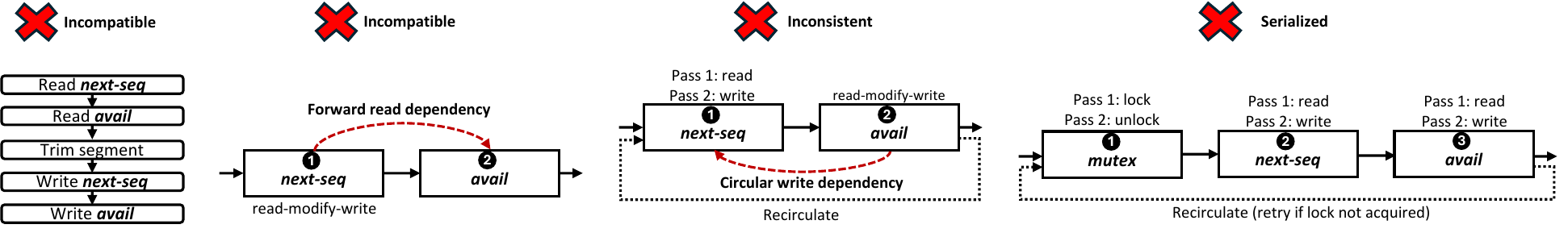}
  \caption{
    Challenges in mapping TCP reassembly onto an
    RMT pipeline.
    \textbf{(a)} CPU\@: the stack freely reads and
    updates state multiple times.
    \textbf{(b)} RMT\@: placing state in different stages introduces a
    \emph{forward read dependency}.
    \textbf{(c)} RMT\@: Deferring updates to resolve this dependency
    creates a \emph{circular write dependency}.
    \textbf{(d)} RMT\@: Serializing updates stalls the pipeline.
  }\label{fig:tcp-dependencies}
\end{figure*}

We make the following contributions:

\begin{itemize}[nosep,leftmargin=*]
  \item \textbf{A match-action TCP design.} We present \sys{}'s design
        principles, showing how to map TCP's dependency-heavy data-path onto a
        unidirectional, match-action pipeline with tight timing and
        resource constraints. While rooted in TCP, these principles apply broadly across stateful
        protocols and programmable network platforms, including SmartNICs,
        FPGAs, and hybrid systems that combine programmable and fixed-function
        components.

  \item \textbf{Prototype on multiple hardware targets.} We demonstrate that
        these principles are realizable on real hardware with a \sys{}
        design and implementation on an Intel Tofino~2
        switch~\cite{tofino2,netberg_aurora810}, showing for the first time that RMT
        can support a TCP stack with ASIC-like performance and energy
        efficiency while retaining programmability. We establish generalizability with our \sys{} port to an FPGA SmartNIC\@; under identical timing
        constraints it supports $3\times$ higher packet rates than
        Tonic~\cite{tonic} and orders of magnitude higher than
        Beehive~\cite{beehive}, while consuming significantly fewer hardware
        resources. Our code is available at
        \url{https://github.com/tcp-acceleration-service/Presto}.

  \item \textbf{Performance, robustness, flexibility, and interoperability.} We
        compare \sys{} to Linux, TAS~\cite{tas}, and RDMA
        (\shortsecref{sec:eval}).\ \sys{} surpasses TAS's peak throughput using
        16 fewer CPU cores and matches RDMA performance and efficiency for
        both RPCs and streaming workloads (up to $25$Mpps per core, enough to saturate
        $1.6$Tbps with $8$K MTU)\@. At scale, \sys{} drives nearly $1$Bpps with
        RPC tail latency near $20\mu s$ and remains resilient to packet
        loss and congestion (up to $2\times$ higher throughput than TAS under
        loss), while interoperating with other TCP stacks.\ \sys{} benefits real applications: it improves a key-value
        store's throughput-per-watt by $2\times$ while maintaining 99.99p tail
        latency below TAS's best case, and enables NVMe-over-TCP to achieve
        RDMA-level performance and CPU efficiency. We demonstrate
        flexibility with transport extensions (Timely~\cite{timely} congestion control, delayed and selective ACKs)
        and application optimizations, including an in-network sequencer for a shared log, similar to NoPaxos~\cite{nopaxos}.
\end{itemize}
 \section{Background}

Can we map TCP's stateful, dependency-heavy data path onto a lock-step
match-action pipeline at line-rate? To approach this question, we first
summarize the RMT programming model that underlies high-speed programmable
switches and SmartNICs (\shortsecref{sec:rmt-architecture}). We then explain
why TCP poses significant challenges for this model
(\shortsecref{sec:rmt-challenges}), as TCP's per-segment processing creates
forward read dependencies (Figure~\ref{fig:tcp-dependencies}b), circular write
dependencies (Figure~\ref{fig:tcp-dependencies}c), and, if we avoid both by
serializing, pipeline stalls (Figure~\ref{fig:tcp-dependencies}d).

\subsection{Match-Action Architecture}\label{sec:rmt-architecture}

The reconfigurable match-action table (RMT) architecture
targets line-rate packet processing. It exploits packet-level parallelism and a
fixed-latency pipeline of stages that apply match-action logic to a per-packet
header vector (PHV). Modern implementations sustain billions of packets per
second with sub-microsecond latency and low ASIC power
consumption~\cite{metamorphosis, netberg_aurora810}.

\paragraph{Pipeline execution.}
Packets enter through a MAC/DMA channel and are parsed into a PHV\@. The PHV
then traverses a sequence of stages that execute in lock-step. Within each
stage, simple ALU operations update PHV fields, accessing \emph{stage-local}
state via tightly-timed memory operations. Because stages are synchronized,
packet processing latency is fixed by the number of occupied pipeline stages.
Implementations typically split processing into ingress and egress pipelines,
with a traffic manager (TM) in between. The TM buffers bursts and schedules
packets to match output bandwidth; it can also participate in lossless
operation via backpressure (e.g., PFC).

RMT provides limited ways to feed information back into the pipeline:
\begin{enumerate*}[label={(\roman*)}]
  \item \textbf{recirculation} (send a packet for another full pass),
  \item \textbf{mirroring} (duplicate PHV and reinject), and
  \item \textbf{packet generator} (emit packets on triggers).
\end{enumerate*} These mechanisms are useful but must be used sparingly:
recirculation doubles latency and consumes pipeline bandwidth; mirroring is
cheaper but still consumes scheduling and pipeline resources.
Finally, uncommon processing and configuration are handled by a CPU control
plane that can directly read and write pipeline state, and exchange packets via
a dedicated DMA channel.

\begin{figure*}
  \centering
\includegraphics[width=\textwidth]{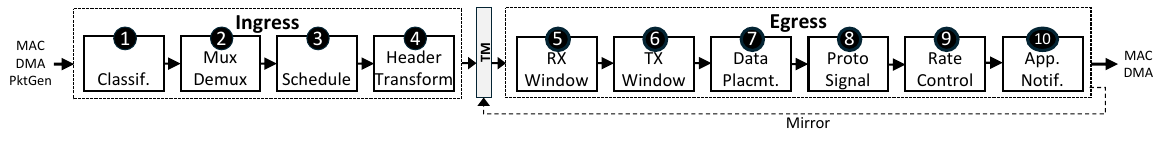}
  \caption{\sys{} RMT data-path functional blocks.}\label{fig:highlevel-arch}
\end{figure*}

\subsection{Challenges for TCP on RMT}\label{sec:rmt-challenges}
TCP's reliable in-order semantics couple each segment's processing to shared
per-connection state (e.g., window boundaries and out-of-order bookkeeping). On
a CPU, the stack can read and update this state multiple times in a single
run-to-completion handler (Figure~\ref{fig:tcp-dependencies}a). On an RMT
pipeline, however, state is stage-local and the PHV flows only forward, so even
simple TCP logic quickly runs into architectural constraints.

Prior work, such as TAS~\cite{tas} and FlexTOE~\cite{flextoe}, splits TCP into
a fast data-path and a slower control path, but assumes a programmable
execution model with run-to-completion processing and flexible memory access
per packet (e.g., a CPU or NPU).\ In contrast, RMT executes a fixed-depth,
lock-step match-action pipeline: packets carry a bounded amount of metadata and
cannot revisit earlier stages, while per-flow state is accessed only through
stage-local stateful operations.\ As a result, the primary challenge on RMT is
not the fast/slow decomposition, but managing TCP's state dependencies without
feedback, buffering, or stalls.

In particular, placing related TCP state---or, more generally, any reliable
transport (\shortsecref{sec:discussion})---in different stages forces one of
three problematic mappings: \begin{enumerate*}[label={(\roman*)}]
  \item if an early stage must update state based on a value stored later, the
        design introduces a \emph{forward read dependency}, which cannot be
        resolved (Figure~\ref{fig:tcp-dependencies}b);
  \item if we defer the update until the later value is known, the update must
        effectively flow backward to earlier-stage state. This creates a
        \emph{circular write dependency} that requires loopback via
        recirculation or mirroring (Figure~\ref{fig:tcp-dependencies}c), which
        consumes bandwidth, adds latency, and can transiently expose stale
        state; and
  \item serializing updates to avoid both (e.g., freezing state and committing
        later) stalls the pipeline and squanders parallelism
        (Figure~\ref{fig:tcp-dependencies}d).
\end{enumerate*}

 \section{\sys{} Overview}\label{sec:design}\label{sec:lock-free-sync}

To overcome the challenges of \shortsecref{sec:rmt-challenges}, \sys{}
rearchitects TCP's per-connection transport logic to maximize RMT pipeline
parallelism. We adopt these design principles, each addressing a specific RMT
constraint:

\begin{enumerate}[label={\arabic*.},nosep,leftmargin=*]

  \item \textbf{Bump-in-the-wire processing (avoid stalls):}
        \sys{} processes each TCP segment at most once in the pipeline (no
        recirculation on the common path) and never buffers segments in the pipeline,
        preserving both bandwidth and low latency under tight memory constraints.

  \item \textbf{Optimistic segment processing (handle forward reads):}
        To deal with forward read dependencies, \sys{} executes the common case
        speculatively (in-window segment receipt and delivery) and defers validation
        to later stages; uncommon cases fall back to corrective actions.

  \item \textbf{Pseudo segment cyclic updates (resolve circular writes):}
        To resolve circular write dependencies without stalling the pipeline, \sys{}
        injects \emph{pseudo segments} via mirroring. These segments are processed
        like regular TCP segments, but are crafted to trigger the required updates to
        earlier-stage state without application-visible side effects.
\end{enumerate}

\noindent
Like many other proposals~\cite{flextoe,tas}, \sys{} accelerates TCP processing
for established connections---the common case for high-performance
applications---and divides the TCP stack into three key components:
\begin{enumerate*}[label={(\roman*)}]
  \item A \emph{control plane} orchestrates connection lifecycle and data-path
        resources, and executes congestion control policies. It may run either
        on the host or on a control CPU alongside the RMT pipeline.
  \item An \emph{application interface} enables direct interaction with the RMT
        data-path, exposing both a POSIX-compatible sockets API (\libsys{}) and
        a zero-copy API (\libsyszc{}) to applications. \ \libsyszc{} allows
        data-intensive applications to access \libsys{} receive and transmit
        buffers directly, avoiding unnecessary data copies, while \libsys{}
        provides a conventional socket API that copies user buffers into
        \libsys{} receive and transmit buffers.
  \item The \emph{RMT data-path} executes core TCP transport logic to ensure
        reliable delivery of ordered byte streams directly to/from \libsys{}.
\end{enumerate*}

\paragraph{Data-path architecture.}
For established connections, \sys{} executes TCP transport logic as a single
RMT pipeline triggered by \emph{events} from three sources: the network MAC
(incoming TCP segments), the host DMA interface (application payload and
notifications), and periodic/explicit triggers (window/credit refresh). These
events induce three logical workflows: RX (network $\rightarrow$ host
delivery), TX (host payload $\rightarrow$ network transmission), and SYNC
(connection state update). Although we describe them separately, they all
traverse the same match-action pipeline and touch shared per-connection state,
reflecting TCP's bidirectional semantics.

To achieve extensibility, \sys{} structures this pipeline as a sequence of
decoupled, independently programmable functional blocks
(Figure~\ref{fig:highlevel-arch}). Each block spans multiple match-action
stages and is subject to the state and dependency constraints of the RMT model.
The \emph{ingress} portion performs stateless classification and preparation:
\Circled{1} Classification identifies whether the event corresponds to RX, TX,
or SYNC and selects the corresponding fast path; \Circled{2} Mux/Demux
retrieves connection and context identifiers; \Circled{3} Scheduler triggers
segment transmissions subject to per-flow rate limits; and \Circled{4} Header
Transformation constructs the required protocol and DMA headers before handing
the packet/metadata to the traffic manager (TM) for egress processing.

The \emph{egress} portion executes the stateful transport logic for all three
workflows. For RX, \Circled{5} Receive Window validates sequence state, detects
gaps/loss signals, and records out-of-order segments, while \Circled{6}
Transmit Window handles acknowledgment processing and loss detection;
\Circled{7} Data Placement translates sequence numbers to host buffer addresses
for DMA\@; and \Circled{10} Application Notification informs the host of newly
arrived data. For TX, \Circled{6} Transmit Window enforces flow control, while
\Circled{9} Rate Control enforces credit-based transmission rates. For SYNC,
the same window and rate control blocks apply credit updates. Finally,
\Circled{8} Protocol \& Congestion Signaling collects congestion/ACK state and
prepares acknowledgments that are emitted without involving the host.

Crucially, \sys{} restricts \emph{mutable} protocol state updates to the
lossless egress pipeline, treating ingress as strictly read-only. This makes
\sys{} \textit{drop resistant}: TM drops are equivalent to network loss and
naturally recovered by TCP retransmission, preserving correctness.

We now describe a baseline instantiation of these blocks that implements a
complete, standards-compliant TCP stack that integrates directly with
real-world datacenter applications running on the host. We focus on segment
reassembly (\shortsecref{sec:reassembly}) and (re-)transmission
(\shortsecref{sec:loss-detection}), as they concentrate the hardest RMT-mapping
challenges and largely determine correctness and throughput. In
\S\ref{sec:rmt-resource-usage}, we evaluate how the same blocks can be reused
and extended to support additional transport extensions and
application-specific optimizations.

\subsection{Segment Reassembly}\label{sec:reassembly}

Segment reassembly involves incorporating any relevant part of an RX segment
within the receive window, delivering in-order data to the application,
buffering out-of-order (OOO) segments for future delivery, collecting
congestion metrics, generating acknowledgments, and notifying \libsys{} of
newly available data. 

\begin{figure}
  \centering
  \includegraphics[width=\columnwidth]{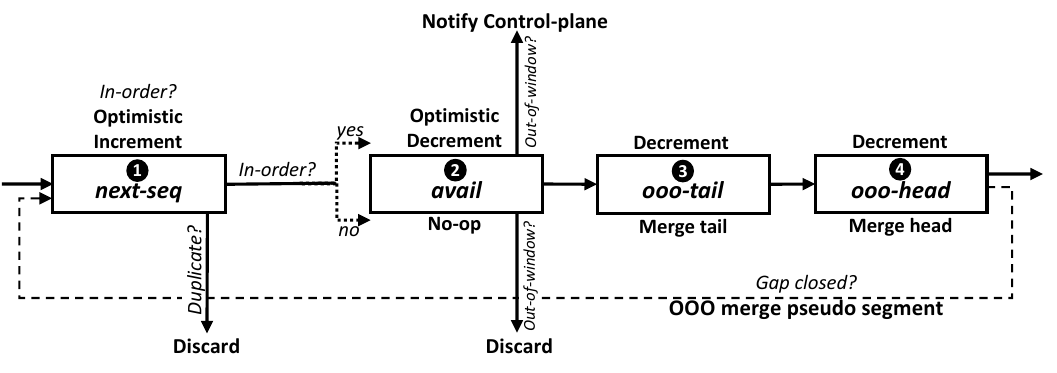}
  \caption{RMT TCP segment reassembly paths in \Circled{5}.}\label{fig:tcp-reassembly-algo}
\end{figure}

\subsubsection{Receive Window Tracking (Figure~\ref{fig:tcp-reassembly-algo})}\label{sec:receive-window}

Dynamic data structures are infeasible in RMT, so \sys{} implements reassembly
using a \emph{fixed} amount of per-connection state. We define the design's
reassembly fidelity by the number of out-of-order (OOO) intervals---contiguous
sequence ranges received ahead of the cumulative acknowledgment point---that it
can track simultaneously. This yields a spectrum from OOO-0 (drop all OOO
segments, no tracking state) to OOO-$n$ (tracking $n$ OOO intervals).
Section~\ref{sec:robustness-eval} quantifies this fidelity-performance
trade-off, while Table~\ref{tab:flex-ext} shows the resource costs.

\paragraph{Baseline reassembly (OOO-0).}
OOO-0 maintains only two state variables: \texttt{next-seq} (the next expected
sequence number) and \texttt{avail} (remaining receive-window space). For an
in-order segment, the receiver must \begin{enumerate*}[label={(\roman*)}]
  \item validate that the segment fits in the advertised window and
  \item advance \texttt{next-seq} and decrement \texttt{avail} by the accepted
        payload length.
\end{enumerate*} Mapping this directly to an RMT
pipeline creates a \emph{forward read dependency}: the stage that advances
\texttt{next-seq} needs the current \texttt{avail} value to validate the
window.

\sys{} resolves this dependency using \emph{optimistic execution}. Well-behaved
senders respect the advertised flow-control window, so \sys{} speculatively
assumes the segment is in-window and validates that assumption later.
Concretely, in \Circled{1}, \sys{} trims/drops duplicates and
\emph{optimistically} advances \texttt{next-seq} for in-order segments,
forwarding a snapshot of the updated \texttt{next-seq} to later stages.
In \Circled{2}, it then decrements \texttt{avail} by the (trimmed)
accepted length and performs the definitive out-of-window (OOW) check using the
now-updated window state.

If the segment (partially) exceeds the advertised receive window (a sender
protocol violation), the speculative update in \Circled{1} has advanced the
receive state incorrectly.\ \sys{} detects this in \Circled{2}, drops the
segment, and raises an exception to the control plane, which restores the
per-connection state to its previous values. During this brief recovery window,
\texttt{avail} may become negative due to the speculative decrement. While
\texttt{avail}<0, subsequent segments fail boundary checks and are dropped, and
\sys{} advertises a zero receive window to pause the sender until the
restoration completes. Because the receiver accepts payload only after final
validation, externally visible behavior stays correct.

\begin{figure}
  \centering
  \includegraphics[width=\columnwidth]{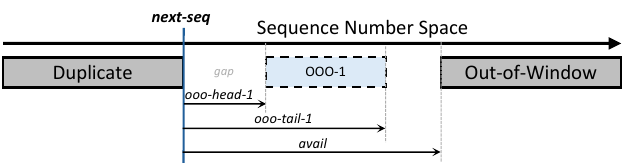}
  \caption{TCP reassembly state (OOO-1) in sequence space.}\label{fig:tcp-reassembly-ooo1-state}
\end{figure}

\paragraph{Robust reassembly (OOO-1).}
OOO-1 improves robustness by tracking a single OOO ``island'' non-contiguous
with \texttt{next-seq}. It adds two per-connection state variables,
\texttt{ooo-tail} and \texttt{ooo-head}, stored as \emph{offsets relative to
\texttt{next-seq}} (Figure~\ref{fig:tcp-reassembly-ooo1-state}). This
representation keeps the OOO interval aligned as the receive window advances.\
An \texttt{ooo-tail}=0 indicates that no OOO interval is currently tracked.

For in-order segments, as \texttt{next-seq} advances, \sys{} decrements
\texttt{ooo-tail} and \texttt{ooo-head} by the same accepted length, preserving
the island's absolute position. For out-of-order segments, if the segment
overlaps with (or is adjacent to) the tracked interval, \sys{} merges it by
updating the interval boundaries. If no interval is active (\texttt{ooo-tail}=0
in \Circled{3}), the arriving OOO segment initializes \texttt{ooo-tail} and
\texttt{ooo-head}. Segments that do not overlap the single tracked range are
dropped.

\begin{figure}
  \centering
  \includegraphics[width=\columnwidth]{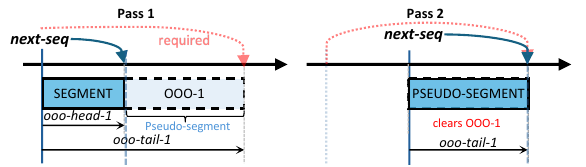}
  \caption{Resolving \emph{circular write dependency} using pseudo-segments (OOO-1).}\label{fig:tcp-reassembly-ooo1-merge}
\end{figure}

A difficult case arises when an in-order segment closes the gap to the tracked
OOO interval, making the byte stream contiguous through the end of that
interval (Figure~\ref{fig:tcp-reassembly-ooo1-merge}). At this point the
receiver should advance the window to the end of the OOO island and free the
corresponding space, i.e., update \texttt{next-seq} and \texttt{avail}.
However, this condition is only known once \texttt{ooo-head} is inspected
(\Circled{4}), while \texttt{next-seq} and \texttt{avail} reside in earlier
stages---a \emph{circular write dependency}.

\sys{} resolves this without stalling the pipeline using a two-step merge. When
\Circled{4} detects that the gap is closed (i.e., \texttt{ooo-head}
reaches 0), it injects a \emph{pseudo-segment}. The pseudo-segment carries no
payload but encodes a sequence range spanning the newly contiguous bytes (from
the current \texttt{next-seq} through \texttt{ooo-tail}). When processed, it
appears as a standard in-order segment, so the OOO-0 logic advances
\texttt{next-seq} and decrements \texttt{avail} and the OOO interval is
cleared.

Intuitively, pseudo-segments allow \sys{} to express backward state updates
using the same forward execution path as real packets, preserving pipeline
invariants without introducing special cases or stalls. Pseudo-segments
preserve correctness during the transient period before the merge commits. If
intervening in-order segments advance \texttt{next-seq} first, the
pseudo-segment's overlapping prefix is naturally trimmed in \Circled{1},
preventing double-accounting. A pseudo-segment merge also cannot create an OOW
condition: the OOO interval was only formed from segments that were previously
validated against the receive window in \Circled{2}. Pseudo-segments are
best-effort; if one is dropped under congestion, the OOO interval remains
tracked and the next triggering segment re-injects it, yielding eventual
consistency.

\paragraph{OOO-$n$.}
The same approach extends to tracking multiple OOO intervals by adding
\texttt{ooo-tail-2}, \texttt{ooo-head-2}, etc.\ across subsequent stages. To
keep processing simple (and avoid sorting primitives), \sys{} maintains two
invariants: \begin{enumerate*}[label={(\roman*)}]
  \item active intervals are stored in increasing sequence order (i.e.,
        \texttt{ooo-tail-$i$} $\ge$ \texttt{ooo-tail-$(i-1)$}).
  \item There are no ``holes''. If \texttt{ooo-tail-$i$}=0 then
        \texttt{ooo-tail-$(i+1)$}=0.
\end{enumerate*}
With these invariants, merging is a single pass: an arriving OOO segment is
tested against each active interval and merged on overlaps. Otherwise, it
cascades to the first inactive slot and initializes a new interval.

\begin{figure}
  \centering
  \includegraphics[width=\columnwidth]{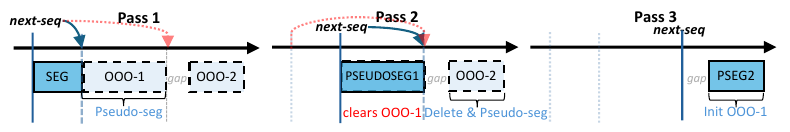}
  \caption{Cascading merge of multiple out-of-order intervals.}\label{fig:tcp-reassembly-ooo-n-merge}
\end{figure}

When an in-order segment closes the gap from \texttt{next-seq} to one or more
intervals, \sys{} again uses pseudo-segments to commit the resulting window
advance. The exceptional case is when closing OOO-1 would leave a hole before
an active OOO-2, violating the second invariant.\ \sys{} handles this with a
cascading merge: in the triggering pass, it snapshots and clears the subsequent
active intervals (Figure~\ref{fig:tcp-reassembly-ooo-n-merge}).
To do so, it treats the in-order segment as having an effective length greater
than the maximum window size for the purposes of processing subsequent OOO
intervals. Since in-order OOO maintenance simply decrements \texttt{ooo-tail-i}
and \texttt{ooo-head-i} by the segment length (as in OOO-1), this
deterministically clears all downstream active intervals with no new
match-action logic.\ \sys{} then injects one pseudo-segment per cleared
interval that includes the snapshots. These pseudo-segments are replayed in
order, advancing the window or re-initializing intervals without leaving holes.

\paragraph{Replenishment.}
As the receiver delivers data, it must replenish window space.\ \libsys{}
triggers the SYNC workflow after the application consumes more than a
configurable fraction of its receive buffer (default: $\frac{1}{4}$). On SYNC,
the Receive Window block increments \texttt{avail} by the amount supplied by
\libsys{}. Because SYNC events traverse the same lossless egress pipeline as
RX, replenishment serializes naturally with segment processing, ensuring
consistent updates.

\subsubsection{Data Placement (\Circled{7} in Figure~\ref{fig:highlevel-arch})} Leveraging our \emph{bump-in-the-wire}
design principle, \sys{} directly places received payload into host memory via
DMA\@, even for OOO segments.
\sys{} maintains a per-connection host receive buffer in
\libsys{}, and the Data Placement block translates TCP sequence numbers into offsets
within this buffer to resolve host memory addresses for DMA transfers.

To handle buffer wraparound, \sys{} double-buffers: the receive buffer is
mapped twice consecutively in \libsys{}'s virtual address space, sharing the
same physical memory. This lets a single contiguous DMA span the logical end of
the buffer rather than splitting into multiple transfers.

\subsubsection{Acknowledgment Generation (\Circled{8} in Figure~\ref{fig:highlevel-arch})}\label{sec:ack-generation} Acknowledgment
generation in \sys{} occurs after segment reassembly. The Protocol \&
Congestion Signaling block governs acknowledgment generation based on
post-reassembly window snapshots.
When an ACK is required, \sys{} generates a pseudo-segment encapsulating the
reassembly state; this segment is mirrored and transformed into a valid TCP
acknowledgment. If the pipeline drops the pseudo segment, the effect is
identical to network packet loss of a TCP acknowledgment. Mirroring
acknowledgments does not increase pipeline workload, as it replaces
host-generated ACK packets rather than introducing additional traffic. As the
TX workflow also traverses the reassembly path, it can generate piggybacked
acknowledgments on outgoing segments based on receive state.\ \sys{} conserves
pipeline bandwidth by coalescing pseudo-segments; for instance, the OOO-1 merge
operation (\shortsecref{sec:receive-window}) piggybacks on the corresponding
ACK pseudo-segment.

\subsubsection{Application Notification (\Circled{10} in Figure~\ref{fig:highlevel-arch})} The Application Notification block informs
\libsys{} when newly received data becomes available for application
consumption, by including a notification inline with the DMA data transfer. For
each \textit{in-order} segment that advances the receive window, \sys{}
notifies \libsys{} of the highest contiguous buffer offset that is ready to be
consumed by the application.

When a received segment closes the gap to an existing out-of-order interval
(\shortsecref{sec:receive-window}), \sys{} issues the notification immediately,
ahead of the subsequent pseudo-segment merge, to minimize notification latency.

\subsection{(Re-)transmission}\label{sec:loss-detection}
(Re-)transmission in \sys{} regulates the sending rate, selects data for
transmission, and recovers from packet loss, driven by data-path transport
state.

\sys{} adopts a \textit{push-based} transmission model optimized for low
latency on uncongested flows, which dominate datacenter workloads~\cite{erpc}. The
data-path grants transmission credits to \libsys{} via SYNC messages; upon
receiving credits, \libsys{} proactively pushes eligible payloads into the
data-path using DMA writes, allowing immediate transmission
aligning with the \emph{bump-in-the-wire} design principle. By contrast,
\textit{pull-based} scheduling adds a roundtrip interaction between \libsys{}
and the data-path, raising latency. The same design supports a pull-based
variant by transforming SYNCs into DMA reads; we leave this to future work.

\subsubsection{Congestion Control}\label{sec:scheduling}
To remain robust to packet loss and congestion, \sys{} supports
DCTCP~\cite{dctcp} by default. Like other TCP stacks~\cite{tas, flextoe, ccp,
zhao2025white, singhvi2025falcon}, \sys{} locates the congestion control policy
in the control plane
for extensibility. The data-path collects per-connection metrics (acknowledged
bytes, ECN-marked bytes, duplicate ACKs) in the Protocol \& Congestion
Signaling block; the control plane periodically collects these metrics and
updates data-path tables with computed transmission rates.

\paragraph{Scheduling via SYNC}
Based on the configured rate, the data-path issues periodic SYNC messages that
convey transmission credits to \libsys{}. SYNCs are generated directly in the
RMT pipeline using the packet generator, without additional control-plane
involvement. The Scheduler block (\Circled{3} in
Figure~\ref{fig:highlevel-arch}) assigns SYNCs to flows according to the chosen
scheduling policy and rate, controlling both SYNC frequency and the credits
granted per SYNC\@. SYNCs also signal retransmission timeouts; in this case,
\libsys{} resets its transmission head and initiates retransmissions.

To be work conserving, \sys{} pauses SYNC workflows for inactive connections
(no data transmission for $N$ RTTs). For uncongested flows with short messages
that underutilize credits, the control plane reduces SYNC frequency.\ \libsys{}
may also self-pace SYNCs for such flows, triggering them only as the transmit
window nears exhaustion.

\subsubsection{Transmission}\label{sec:tx}
\libsys{} pushes new data from the host transmit buffer into the data-path via
DMA when sufficient credits are available. TCP sequence numbers are derived
from DMA offset within the transmit buffer. The Transmit Window block tracks
transmit state using logic analogous to the receive side
(\shortsecref{sec:receive-window}): segments are validated against the current
transmit window. Segments below the cumulative acknowledgment point are dropped
and segments extending beyond the window advance it accordingly (\Circled{3}).
Transmission is credit-limited and flow-controlled in the Rate Control block
(\Circled{9}), ensuring the sending rate respects both allocated credits and
the peer's advertised receive window.

\subsubsection{Acknowledgment Processing}\label{sec:ack_rx}
For received ACKs, the Transmit Window block (\Circled{6} in
Figure~\ref{fig:highlevel-arch}) updates transmit window state, including the
cumulative acknowledgment point, merging any SACK blocks to track lost
segments, mirroring the logic of the receive side
(\shortsecref{sec:receive-window}). When sufficient transmit window space
becomes available, the Application Notification block (\Circled{10}) transforms
the ACK into a notification to \libsys{} to reclaim buffer space.

\paragraph{Fast Retransmission}
Upon three duplicate ACKs, \sys{} triggers fast retransmission and notifies
\libsys{} of the updated transmit window state (including SACK blocks) via
DMA\@. The Rate Control block simultaneously reduces transmission credits,
similar to TCP Reno, and conveys the updated credits to \libsys{} in the same
DMA notification.

 \section{Implementation}\label{sec:impl}
\sys{}'s data-path is implemented on an Intel Tofino~2-based switch,
a commercially available RMT pipeline architecture.
The system comprises 20,775 lines of code (LoC) across the data-path,
control-plane, and \libsys{}.  The P4 code for the offloaded data-path has
2,597 LoC. The control plane is split between the host and
a switch-resident CPU (\texttt{SwitchConf}), which configures the match-action
tables, executes congestion control policy, and processes control notifications.
\texttt{SwitchConf} interacts with the data-path via Intel's proprietary bfrt
C++ API\@. They are implemented in 3,352 and 6,917 C++ LoC respectively.
\ \libsys{} is implemented in 7,909 C++ LoC.
For DMA, \libsys{} utilizes RDMA\@ (\texttt{libibverbs}) without kernel
intervention. The host control-plane and POSIX emulation in \libsys{} are
adapted from TAS\@.
Our prototype leverages some RDMA and Tofino~2 features for optimizations,
described below.

\paragraph{DMA}
\libsys{} interacts with the data-path via RDMA unreliable connection (UC)
queue pairs, one per application context (thread), akin to per-core
NIC queues, with each context managing multiple connections. Data transfers
use RDMA Writes (with immediate for notifications) directly to/from
per-connection payload buffers.\ \libsys{} allocates per-connection
payload buffers and registers them with the RDMA NIC, storing the buffer's
virtual address and remote memory key in the data-path.
Based on available transmission credits, \libsys{} can issue RDMA writes larger
than the MTU, relying on the NIC's segmentation offload to dynamically packetize
the data, and paces transmissions over the SYNC interval to smooth bursts.
Note that we use RDMA solely for DMA---reliability, ordering, and
congestion control are all handled by the data-path. For flow control, PFC is
enabled between the host and the switch. External-facing switch ports connected
to hosts outside the subnet using TCP do not require PFC\@. For \sys{}
implementations on RMT-based SmartNICs with PCIe-based DMA interfaces, PFC is
unnecessary for lossless operation.

\paragraph{Interrupt-driven operation}
By default, \libsys{} polls the completion queue for events. After 10ms of
inactivity, it switches to interrupt-mode by via RDMA completion event
channels,
letting application sleep while waiting for IO\@. Periodic SYNC notifications
from the data-path are also paused for inactive connections. This reduces the
excessive CPU overhead and energy consumption due to polling.

\subsection{Testbed Constraints}\label{sec:testbed-constraints}
Below are some constraints of our testbed, which are specific to the Tofino~2
architecture and can be elided in non-switch platforms.

\paragraph{\sys{}-\sys{} connections}
Same-switch \sys{}-\sys{} connections require recirculation because
both server and client stacks reside on one switch; non-switch
platforms avoid this limitation. All TCP traffic for such connections,
segments and ACKs alike, must recirculate to the data-path. For example, an RPC
request leaves \libsys{} as an RDMA write, the switch data-path converts it to a
TCP segment, recirculates it for peer reassembly, and converts it back to an
RDMA write. The RPC response and ACKs also undergo recirculation. Recirculation
adds latency ($\approx 750ns$) and is limited by $3 \times 100$G
recirculation port bandwidth, with connections sharded across the three ports.

\paragraph{Single-pipeline operation}
Although the Tofino~2 features multiple pipelines, our prototype confines
data-path processing to a single pipeline. Because we maintain all mutable
state within the egress pipeline to ensure drop resistance
(\shortsecref{sec:design}), the sender and receiver must share the same
pipeline, precluding cross-pipeline connections. The remaining pipelines can
operate independently as isolated subnets.

\paragraph{Checksums}
Our prototype omits payload checksums (including TCP and RDMA iCRC), as the
Tofino~2 checksum engine lacks support; a production version would include
hardware checksum units. Because the RDMA iCRC is appended after the payload,
our prototype requires the sender to add Ethernet padding for the iCRC in TCP
segments.

 \section{Evaluation}\label{sec:eval}

We evaluate \sys{} by addressing the following questions:
\begin{itemize}[leftmargin=*,nosep]
  \item \textbf{Performance:} Can \sys{} deliver predictable $\mu$s
        latency and high message rate for RPCs
        (\shortsecref{sec:rpc-eval}), sustain terabit throughput
        with near-zero CPU utilization (\shortsecref{sec:stream-eval}), scale
        efficiently to thousands of applications and connections
        (\shortsecref{sec:rpc-eval}), and benefit real-world
        storage (\shortsecref{sec:storage-eval}) and key-value stores
        (\shortsecref{sec:energy-efficiency})?
  \item \textbf{Efficiency:} Does \sys{} improve performance-per-watt
        over host-based stacks (\shortsecref{sec:energy-efficiency})?
  \item \textbf{Flexibility:} Can \sys{} customize transport logic
        for critical datacenter applications (\shortsecref{sec:rmt-resource-usage})?
  \item \textbf{Robustness:} Is \sys{} resilient to loss and
        congestion, and does it ensure performance
        isolation between latency-sensitive and bandwidth-intensive
        workloads (\shortsecref{sec:robustness-eval})?
  \item \textbf{Generalizability:} Do \sys{}'s pipelined transport logic
        principles extend to FPGA-based SmartNICs (\shortsecref{sec:eval-fpga})?
\end{itemize}

\paragraph{Testbed}
Our evaluation cluster comprises eight Intel Xeon Gold 6430 machines, each with
32-core (64 hyperthreads) running at 2GHz, 256GB RAM, and 126.5MB aggregate
cache. Each machine is equipped with a RDMA-capable 200G Mellanox ConnectX-6
NIC, all connected to an 32$\times$400G Intel Tofino-2 based Netberg Aurora 810
switch. We configure the NICs in 100G mode because the switch fails to
negotiate with the NICs at 200G speeds. Two machines are designated as servers,
with Priority Flow Control (PFC) enabled for lossless RDMA, while the remaining
machines function as clients. To ensure high performance and low latency, we
disable turbo boost and set frequency scaling to performance mode. Power
measurements are obtained via the IPMI interface of both the machines and the
switch.\ \texttt{SwitchConf} runs on a 4-core (8 hyperthreads) onboard Intel
Xeon D-2123IT CPU\@, and its power consumption is included in the switch power
measurements.

\paragraph{Baselines}
We compare \sys{} performance against the Linux TCP stack, the TAS
kernel-bypass TCP stack, and RDMA\@. \sys{} interoperates with both TCP
baselines (cf.~\shortsecref{sec:testbed-constraints}).
Unless otherwise specified, clients use the TAS network stack to generate
network traffic. TAS struggles with large flows due to high payload copy
overheads, failing to saturate the 100G line rate with a single connection. To
mitigate this bottleneck in such client workloads, we elide payload copying in
TAS and transmit junk data instead (TAS-nocopy). Across all TCP-based
baselines, we use identical application binaries built on a sockets-based
\texttt{epoll()} interface; for the RDMA baseline, we adapt the benchmarks to
use the \texttt{libibverbs} interface (a separate verbs-based code path),
whereas \sys{} uses the same sockets-based applications via its POSIX
interposition layer. We confirm real-world application compatibility through
FlexKVS (\shortsecref{sec:energy-efficiency}) and SPDK NVMe-oF
(\shortsecref{sec:storage-eval}); more broadly, our POSIX emulation layer is
\texttt{epoll()} and \texttt{libevent} compatible, and runs unmodified
datacenter applications, including \emph{memcached}~\cite{memcached},
\emph{nginx}~\cite{nginx}, and NVMe-oF clients
(cf.~\cite{flextoe,stolet2023virtuoso}, \shortsecref{sec:storage-eval}).
Unless otherwise noted, our evaluation configures \sys{} with OOO-1 reassembly
and a simple OOO-0 (go-back-N) sender recovery as a baseline. We also
implemented OOO-1 SACK, which enables selective recovery using the same OOO
interval abstraction, but compilation issues in the Intel Tofino~2 toolchain
prevent hardware evaluation. We therefore evaluate OOO-1 SACK in simulation
(Section~\ref{sec:robustness-eval}).
Importantly, go-back-N is not fundamental to \sys{}'s pipeline architecture: it
is a baseline choice that offers the best performance-complexity trade-off, and
\sys{} supports higher-fidelity reassembly and selective recovery mechanisms
(cf.~\shortsecref{sec:robustness-eval}, Table~\ref{tab:flex-ext}).
We use DCTCP as the default congestion control policy.

\subsection{Remote Procedure Calls (RPCs)}\label{sec:rpc-eval}
RPC workloads demand low latency, high packet rates, and scalability across
many applications and connections, requirements that heavily tax host TCP
stacks.\ \sys{}'s bump-in-the-wire design efficiently delivers constant, low
TCP processing latency even at multi-million segment rates, fully eliminating
all CPU overhead. We demonstrate the benefits with an echo server benchmark.

\begin{figure}[t!]
  \centering
  \includegraphics[width=\linewidth]{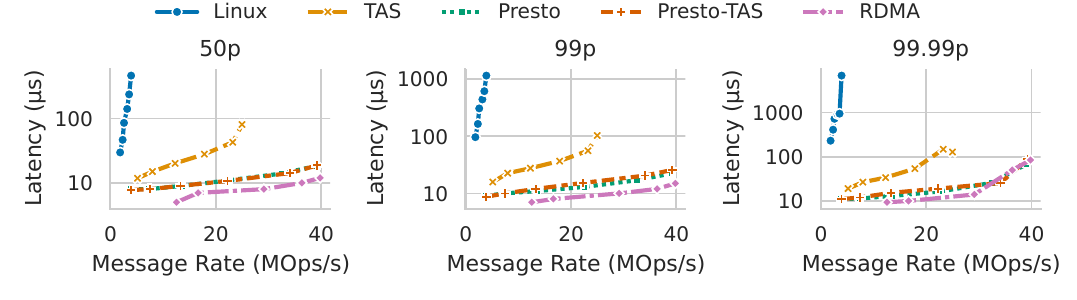}
  \caption{Load-latency across network stacks.
}\label{fig:rpc-load-latency}
\end{figure}

\subsubsection{Load-latency}
Figure~\ref{fig:rpc-load-latency} presents the round-trip latency profile of a
32-core echo server across stacks, as load increases. The benchmark scales
connections across multiple clients, each leaving a single 64B RPC in-flight.
\sys{}'s performance closely matches RDMA\@, sustaining nearly $40$ MOps/sec
($1.67\times$ TAS) with an $18\mu s$/$24\mu s$ median/99.99p tail latency.
Recirculation in \sys{}-\sys{} connections adds $\sim 2\mu s$ relative to
RDMA\@ (\shortsecref{sec:impl}). Our setup bypasses egress processing for
non-\sys{} packets, reducing baseline RTTs; a conventional switch would require
both ingress and egress. Both \sys{} and RDMA experience 99.99p tail latency
spikes as they approach the NIC's processing limits. A hybrid setup (\sys{}
server with TAS clients) achieves a similar peak throughput but with slightly
higher latency due to client-side TAS overheads.
Linux is especially inefficient, delivering $10\times$ lower throughput and
$78\times$ higher latency than \sys{}.\ \sys{}'s sockets interface adds a copy
between \libsys{} and user buffers; \libsyszc{} eliminates it, reaching RDMA
parity.

\begin{figure}
  \centering
  \begin{subfigure}[]{0.66\columnwidth}
    \centering
    \includegraphics[width=\columnwidth]{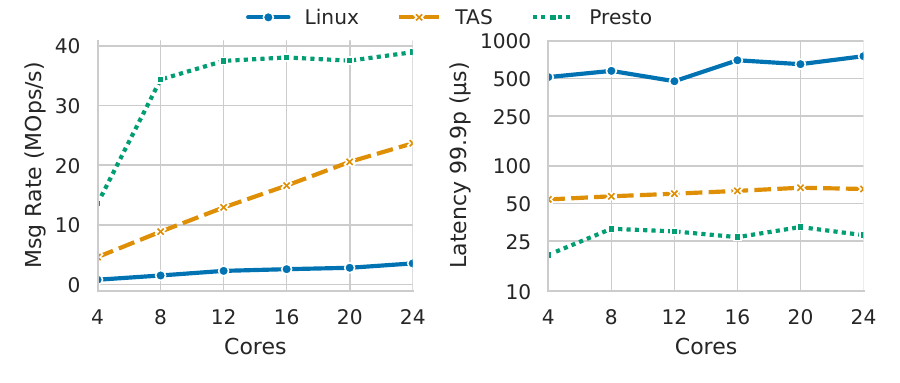}
\end{subfigure}
  \hfill
  \begin{subfigure}[]{0.33\columnwidth}
    \centering
    \includegraphics[width=\columnwidth]{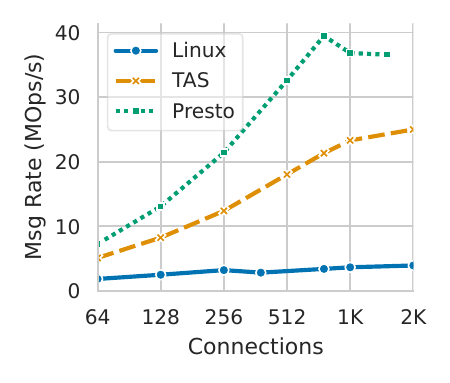}
\end{subfigure}
  \caption{Core and connection scalability.}\label{fig:rpc-core-scalability}\label{fig:rpc-conn-scalability}
\end{figure}

\subsubsection{Scalability}
We evaluate scalability along cores, connections, and pipeline capacity. We
repeat the previous experiment sweeping across host, core, and connection
counts, finding the throughput-latency knee for each configuration.

\paragraph{Cores}
Figure~\ref{fig:rpc-core-scalability} shows that \sys{} scales linearly with
server cores. Data-path performance is independent of core count, as \sys{}
multiplexes connections onto per-core DMA channels (\textit{contexts}),
steering payload directly to application cores (like aRFS~\cite{aRFS}) and
avoiding RDMA scalability limits. Performance peaks at 8 cores, where \sys{}
hits the NIC line rate, surpassing TAS's peak throughput while keeping tail
latency $4.3\times$ lower. Linux remains far behind.

\paragraph{Connections}
Figure~\ref{fig:rpc-conn-scalability} showcases \sys{}'s superior connection
scalability for a single host.\ \sys{} saturates NIC bandwidth at 768
connections, achieving $1.8\times$ TAS and $10\times$ Linux throughput. Beyond
NIC line rate, additional connections slightly reduce throughput.
Accessing connection state via stage-local memory at constant latency, \sys{}
sustains 32K connections without performance loss.
Sharding TCP state tables across stages or multiple pipelines supports higher
connection counts (\shortsecref{sec:rmt-resource-usage}). Other
high-performance stacks degrade with many connections due to cache
pressure~\cite{erpc} or limited on-NIC memory~\cite{flextoe}.

\begin{figure}
  \centering
  \begin{subfigure}[]{0.49\columnwidth}
    \centering
    \includegraphics[width=\columnwidth]{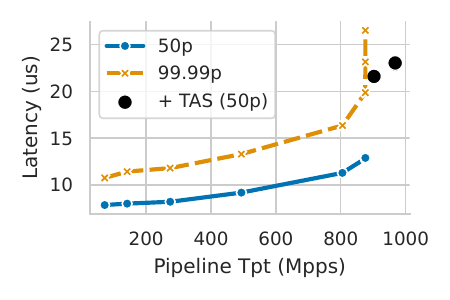}
\end{subfigure}
  \hfill
  \begin{subfigure}[]{0.49\columnwidth}
    \centering
    \includegraphics[width=\columnwidth]{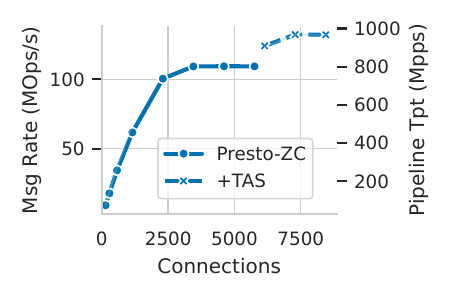}
\end{subfigure}
  \caption{Pipeline packet processing scalability.}\label{fig:pipeline-stress}
\end{figure}

\paragraph{Pipeline}
Each RMT pipeline in our testbed is capable of processing $1.2$ Bpps across
$8\times 400$G links. Using 8 hosts running \sys{}, exchanging 8B echo RPCs,
stresses the pipeline. Each RPC generates 8 packets through the egress
pipeline---4 for the request (TX, RX, ACK generation, ACK processing) and 4 for
the response. Figure~\ref{fig:pipeline-stress} highlights pipeline scalability.
Latency stays flat and narrow up to $850$ Mpps ($71\%$ utilization), after
which recirculation bandwidth saturates and tail latency rises. Two more TAS
clients (black dots) push throughput near $1$ Bpps ($80\%$ utilization), with
aggregate pipeline throughput nearing $500$ Gbps ($130$ MOps/sec goodput) even
at 8B payloads. Latency is higher in this configuration due to client-side TAS
overheads. We have tested scalability up to 8K concurrent connections and 256
contexts in this benchmark. Notably, \sys{} does not experience ACK
pseudo-segment drops, even under such extreme loads; the mirroring engine
reliably sustains the required packet rate ($\approx 250$ Mpps). Because OOO
pseudo-segments are piggybacked on ACKs (\shortsecref{sec:ack-generation}),
\sys{} ensures robust OOO reassembly and recovery without imposing additional
load on the pipeline.

\subsubsection{Energy Efficiency}\label{sec:energy-efficiency}
We measure the power consumption of the TCP stacks using
FlexKVS~\cite{flexnic}, a scalable, high-performance key-value store inspired
by memcached. Short RPCs dominate these workloads, with host TCP stacks
consuming up to 48\% of per-request CPU cycles~\cite{flextoe}.
Our benchmark uses 1M uniformly distributed key-value pairs (8B keys, 64B
values, $0.99$ GET:SET ratio). Baseline power is first recorded for the server
(including NIC) and switch. The workload then runs for 5 minutes while
measuring total power and application performance. We vary server threads and
connections, with each connection limited to a single in-flight request.

\begin{figure}
  \centering
  \begin{subfigure}[]{0.49\columnwidth}
    \centering
    \includegraphics[width=\columnwidth]{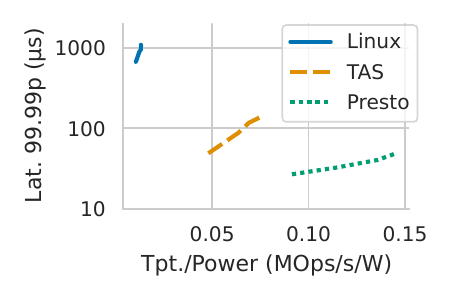}
\end{subfigure}
  \hfill
  \begin{subfigure}[]{0.49\columnwidth}
    \centering
    \includegraphics[width=\columnwidth]{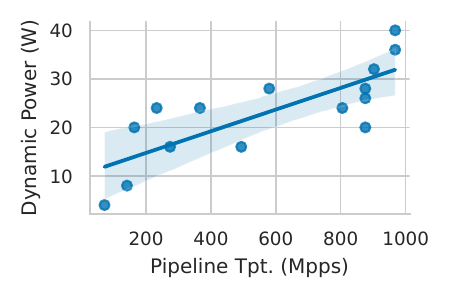}
\end{subfigure}
  \caption{\textbf{(a)} Throughput-per-watt and tail latency for FlexKVS key-value store.\ \textbf{(b)} Pipeline
    dynamic power consumption with utilization.}\label{fig:energy-efficiency}\label{fig:flexkvs-power}\label{fig:pipeline-power}
\end{figure}

Figure~\ref{fig:flexkvs-power}a shows throughput-per-watt and tail latency
across server configurations (\# threads, \# connections).\ \sys{} consistently
beats host stacks: even its worst case exceeds TAS's best by $1.2\times$ in
throughput-per-watt at $5\times$ lower 99.99p latency; its best configuration
doubles TAS's throughput-per-watt and beats TAS's best tail latency. Linux
fares far worse, with at least an order of magnitude lower efficiency and two
orders higher latency.
Switch power varies by no more than 4W between baselines and \sys{}, with
efficiency driven primarily by reduced host CPU usage. RMT-based switches
consume power comparable to non-programmable commodity switches of similar
bandwidth~\cite{agrawal2020intel,metamorphosis}, in contrast to power-hungry
FPGAs~\cite{amdalveou200}. 

Finally, Figure~\ref{fig:pipeline-power}b breaks down dynamic power (increase
from idle, measured over 10 minutes) as a function of pipeline throughput,
showing \sys{} processes TCP at $\approx 25$ Mpps/W for 8B payload packets.
These switch measurements include the on-board CPU, RMT pipeline, transceivers,
and other fixed-function components (cf.\
Table~\ref{tab:tofino-resource-usage}).

\subsection{Byte Streaming}\label{sec:stream-eval}

Streaming bandwidth is critical for modern storage and AI workloads.
CPU-based stacks like Linux and TAS fail to saturate line rate on a single
core, even with hardware-assisted payload copy
optimizations~\cite{cai2021understanding,flextoe}. Specialized DMA copy
engines~\cite{zeronic} do not resolve this, as single-core protocol processing
is the bottleneck (\shortsecref{sec:related}). Consequently, hardware
transports like RDMA and Chelsio TOE, despite their inflexibility, are
increasingly preferred for
streaming~\cite{bai2023empowering,gao2021cloud,chelsio,qian2024alibaba}. Can
\sys{} match their performance?

\subsubsection{Single-core Throughput}
In this benchmark, a client running TAS-nocopy transmits packets in open-loop
1MB streaming bursts (similar to iperf~\cite{iperf}) to a single server core
that discards incoming payloads. To provide different intensities of protocol
to payload processing, the benchmark varies the segment size by adjusting the
maximum transmission unit (MTU).
To avoid payload copying in the sockets layer, we use the \libsyszc{}
interface, comparing against the equivalent low-level TAS interface (TAS-ll),
though TAS still requires copying the payload between network buffers and
application memory.

\begin{figure}
  \centering
  \includegraphics[width=\columnwidth]{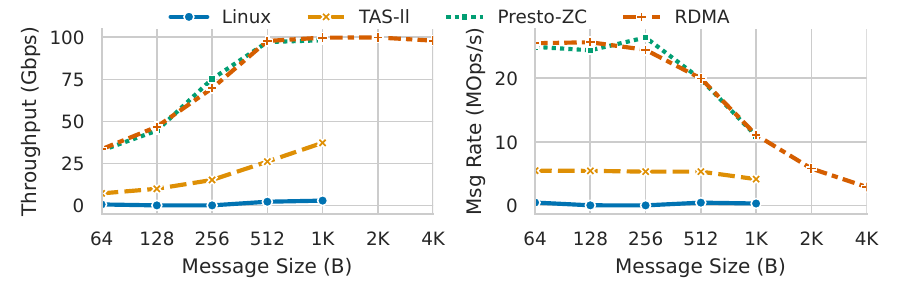}
  \caption{Streaming performance across segment sizes.}\label{fig:stream-bandwidth}
\end{figure}

Figure~\ref{fig:stream-bandwidth} shows that \sys{} matches RDMA's single-core
packet rate of \textasciitilde{}25\,Mpps, independent of packet size, reaching
line rate for higher MTUs, sufficient to exceed 1.6\,Tbps ($= 25M \times 8
\times 8K = 1600G$) with 8K MTU\@. The bottleneck is NIC and CPU
processing~\cite{kalia2016design}, not the RMT pipeline, whose capacity is far
greater (cf. Figure~\ref{fig:pipeline-stress}). Even a single connection can
exploit the pipeline's full parallelism and bandwidth.
TAS is limited to \textasciitilde{}5\,Mpps per core---a bottleneck shared by
NPU-based stacks~\cite{flextoe}, dropping further above 512\,B due to copy
overhead. Linux, again, performs poorly.

\subsubsection{NVMe-over-TCP}\label{sec:storage-eval}
Modern datacenters are rapidly adopting large-scale disaggregated storage.
We integrate SPDK~\cite{spdk}, an open-source user-space storage stack, with
\sys{} to show that NVMe-oTCP can rival RDMA in both performance and
efficiency, while retaining a TCP/POSIX programming model.

We extend SPDK with a new transport that interfaces with \sys{} via
\libsyszc{}, delivering NVMe commands and payloads into per-connection buffers
directly accessible by SPDK, avoiding intermediate copies and enabling high
throughput. A key challenge in supporting NVMe-oTCP is that NVMe devices
exploit parallelism by allowing commands to complete out-of-order (OOO). To
support this, we provide a modified version of \libsyszc{} that exposes
explicit mechanisms to free connection buffer space out-of-order while tracking
the buffer head. NVMe-oRDMA provides similar functionality by requiring SPDK to
post multiple receive buffers to the RDMA work queue, which are freed
independently upon command completion.

We evaluate our NVMe-oTCP storage target using 4KB random writes. In our setup
with a single NVMe SSD, \sys{} can easily exceed the 1M IOPS disk bandwidth,
even when only partially utilizing a server core. To shift the bottleneck from
the device to the CPU, we configure a RAM block device~\cite{spdk_rambdev},
which emulates Non-Volatile Memory (NVM) or Compute eXpress Link (CXL)-attached
memory.

\begin{figure}
  \centering
  \includegraphics[width=\columnwidth]{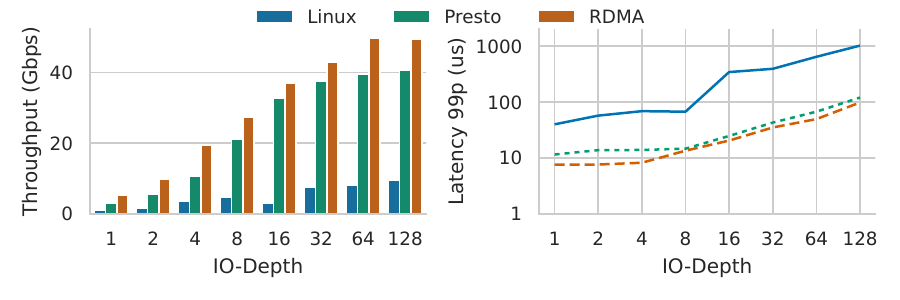}
  \caption{SPDK NVMe-oF performance: 4K Random Writes.}\label{fig:spdk-perf}
\end{figure}

Figure~\ref{fig:spdk-perf} compares the performance of a single-core SPDK
NVMe-oF target using the Linux kernel stack, the \sys{} transport, and RDMA\@,
as IO-depth increases. We use the SPDK perf~\cite{spdk_nvme_perf} tool on a
remote client node\footnote{SPDK perf can generate more load compared to
fio~\cite{spdk_nvme_perf}.} that drives the load through \sys{}'s POSIX sockets
interposition layer. We use an MTU of 4,400 with Linux, 4,096 with RDMA and
\sys{}.

\sys{} closely matches RDMA in throughput and tail latency until the server
core saturates at high IO depths from storage-stack processing and RAM block device
copies.\ \sys{}'s peak throughput is slightly lower from
periodic SYNC grant overhead, and its latency slightly higher due
to client-side sockets layer copies. Linux, by contrast, incurs
extra copies and higher stack overhead, leaving fewer application cycles and
$> 4\times$ lower throughput.

\paragraph{Takeaway}
Matching RDMA in this end-to-end storage setting shows that TCP semantics need
not imply CPU-bound transport processing: \sys{} achieves RDMA-class efficiency
while preserving TCP/POSIX compatibility and leaving room for transport
evolution in the data-path.

\subsection{Flexibility}\label{sec:rmt-resource-usage}
Unlike fixed-function hardware transports, \sys{} provides programmable
transport functionality within the constraints of RMT pipelines, enabling
customization both within and beyond standard TCP features, stack design, and
application integration. We evaluate flexibility systematically: we first
quantify baseline resource usage and then measure the incremental cost of
extensions along key design dimensions.

\paragraph{Resource footprint}
We begin by quantifying the resource footprint of our baseline design. As shown
in Table~\ref{tab:tofino-resource-usage}, it consumes a fraction of available
pipeline resources while supporting 32K connections and 1K contexts.\ \sys{}'s
data-path sustains 3.2 Tbps ($8 \times 400$G) line-rate or 1.2 Bpps at a total
processing latency (excluding queuing delays) of 395ns and a power consumption
of 2.81W for the RMT pipeline (excluding SerDes, MAC, TM). With basic L2/L3
forwarding implemented, the low footprint leaves ample headroom for new
functions. For manageability, \sys{} exports telemetry and debugging metrics
for internal transport state via P4 counters---OOO segments, retransmissions,
congestion events, and pseudo-segments.

\paragraph{Coverage of the design surface}
Transport evolution is broad, but most changes fall into a small number of
axes: \begin{enumerate*}[label=(\roman*)]
  \item transport semantics (e.g., ACK and reassembly strategy, loss recovery
        hints),
  \item congestion control hooks (which metrics are measured, and how they
        drive pacing/scheduling), and
  \item application integration (what events are exposed and how data is
        placed/copied).
\end{enumerate*}
\sys{}'s modular blocks expose these axes directly, so extensions typically require
localized changes rather than redesigning the pipeline. Conversely,
\sys{} does not aim to support unbounded per-flow data structures or complex
control-flow in the data-path; such functionality is delegated to the control
plane or endpoints.

\begin{table}[t!]
  \small
  \begin{tabular}{lr}
    \textbf{Resource}                        & \textbf{Usage} \\
    \toprule
    Stages                                   & 14/20          \\
    Processing Latency                       & 395 ns         \\
    MAU Power (\textit{worst case} per-pipe) & 2.8 W          \\
    \midrule
    \textbf{Memory}                          &                \\
    SRAM                                     & 19.8 \%        \\
    Map RAM                                  & 14.5 \%        \\
    TCAM                                     & 0.4  \%        \\
    \midrule
    \textbf{Compute}                         &                \\
    VLIW Instruction                         & 9.5 \%         \\
    Stateful ALU                             & 26.3 \%        \\
    Match Crossbar                           & 10.8 \%        \\
    Gateway                                  & 26.6 \%        \\
    \bottomrule
  \end{tabular}
  \caption{Tofino~2 RMT pipeline resource usage for \sys{}.}\label{tab:tofino-resource-usage}
\end{table}

\subsubsection{Case Studies}\label{sec:flex-case-studies}
\sys{}'s architecture is modular (Figure~\ref{fig:highlevel-arch}), with each
block exposing a distinct dimension of programmability. These dimensions
include: transport semantics (reassembly and ACK strategies),
congestion control (policy, metrics, and scheduling), and application
integration (notifications, APIs, and data placement).
We next illustrate how these flexibility dimensions can be exercised in
practice through multiple extensions and reassembly strategies, each time
incurring a modest resource cost (Table~\ref{tab:flex-ext}).

\paragraph{Reassembly fidelity} Our reassembly fidelity spectrum (\shortsecref{sec:receive-window}) enables
operators to trade-off resource usage for robustness by tuning the Receive
Window block. As shown in Table~\ref{tab:flex-ext}, OOO-$n$ reassembly variants
fit within Tofino~2 constraints; each increment in fidelity requires two
additional stages and results in a modest power increase.

\paragraph{Transport extensions}
We modify the Protocol \& Congestion Signaling block to support alternative
acknowledgment strategies, including delayed and selective ACK variants.

Delayed or cumulative ACKs~\cite{rfc1122} acknowledge every $M$ consecutively
received segments, reducing sender and network load~\cite{dctcp}. For instance,
for RPCs, piggybacking the ACK onto the response can halve sender overhead. In
\sys{}, a counter tracks unacknowledged segments, issuing an ACK pseudo-segment
at threshold $M$ (or immediately for OOO segments); transmitted segments reset
it. The control plane manages the delayed ACK timer. Benchmarking with a \sys{}
echo-server and TAS client (single fast-path core) shows delayed ACKs improve
peak throughput by 24\% by lowering client CPU load.

Selective ACKs~\cite{rfc2018} enable targeted retransmissions of lost segments,
accelerating recovery in high-loss scenarios. We implement 1-OOO selective ACKs
by extending the ACK generation logic to encode the 1-OOO interval, as
discussed in \shortsecref{sec:receive-window}, and verify that the design fits
within Tofino~2 constraints (cf.~\shortsecref{sec:robustness-eval}).

\paragraph{Congestion control}
Congestion control remains an evolving research area with numerous proposals
addressing diverse network conditions and performance objectives~\cite{1rma}.
Timely~\cite{timely}, which infers congestion from round-trip time (RTT)
variations, is a notable example. Recent RMT hardware
enhancements~\cite{FPISA,tofino2}, including support for lightweight arithmetic
and look-up table approximations, enable more sophisticated computations within
the data-path. As a concrete example, we implement Timely by leveraging
Tofino~2's hardware timestamping to add TCP timestamps~\cite{rfc7323} and
compute per-flow RTT exponentially weighted moving averages (EWMAs) in the
data-path using its low-pass filter unit~\cite{tofino2_ewma}. The control plane
periodically adjusts transmission rates based on these RTTs, while the
data-path enforces the rate limits.

\paragraph{Application co-design}
Databases~\cite{delos}, storage clusters~\cite{corfu}, and replicated state
machines~\cite{scalog} rely on shared logs to linearize state
updates~\cite{luo2024lazylog,corfu}. Inspired by NoPaxos~\cite{nopaxos}, we
accelerate a shared log microbenchmark with a tailored \sys{} API\@: the
application registers its log buffer and client connections, and \sys{}
sequences incoming records across connections, appending them directly to the
log via a buffer tail pointer within the data-path~\footnote{Clients ensure log
records are not fragmented across TCP segments.}. Our design guarantees that
records from the same connection are appended to the shared log in order,
resulting in a linearized shared log. If OOO segments arrive, \sys{} reserves
space up to the highest sequence number, creating gaps that the application
later fills from the corresponding segments \sys{} redirects to it.

We modify the Data Placement, Application Notification, and the \libsys{} API
layers to accelerate a shared log microbenchmark in this way. With up to 32
clients (64--1024B records, 256 in-flight per client), \sys{} saturates
line-rate with 1 core, versus 5 for TAS (3 TAS + 2 app). The gain comes from
removing sequencing, protocol handling, and payload copying from the CPU in the
common case (no loss)\@. Similar RDMA acceleration is hard, especially for
variable-sized records, since it needs multiple round-trips to sequence and
write each record~\cite{kalia2016design}.

\begin{table}[t!]
  \small
  \begin{tabular}{lcrr}
    \textbf{Extension} & \textbf{Blocks (Fig.~\ref{fig:highlevel-arch})} & \textbf{P4 LoC} & \textbf{Power (W)} \\
    \toprule
    \multicolumn{4}{l}{\textbf{Reassembly Fidelity}}                                                            \\
    OOO-0              & \Circled{5}                                     & 176             & 0.19               \\
    OOO-1              & \Circled{5}                                     & 257             & 0.35               \\
    OOO-2              & \Circled{5}                                     & 343             & 0.41               \\
    \midrule
    \multicolumn{4}{l}{\textbf{Transport Ext.}}                                                                 \\
    Delayed ACKs       & \Circled{8}, Control Plane                      & 42              & 0.05               \\
    1-OOO SACK         & \Circled{8}, \Circled{4}                        & 152             & 0.12               \\
    \midrule
    \multicolumn{4}{l}{\textbf{Congestion Control}}                                                             \\
    Timely             & \Circled{8}, Control Plane                      & 64              & 0.37               \\
    \midrule
    \multicolumn{4}{l}{\textbf{Application Co-design}}                                                          \\
    Shared log         & \Circled{7}, \Circled{10}, \libsys{}            & 156             & 0.29               \\
    \bottomrule
  \end{tabular}
  \caption{Marginal cost of \sys{} extensions.}\label{tab:flex-ext}
\end{table}

\subsection{Robustness}\label{sec:robustness-eval}

\begin{figure}
  \centering
  \begin{subfigure}[]{0.49\columnwidth}
    \centering
    \includegraphics[width=\columnwidth]{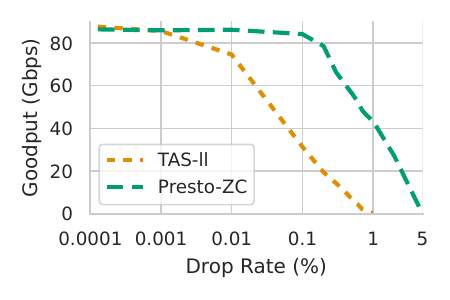}
  \end{subfigure}
  \hfill
  \begin{subfigure}[]{0.49\columnwidth}
    \centering
    \includegraphics[width=\columnwidth]{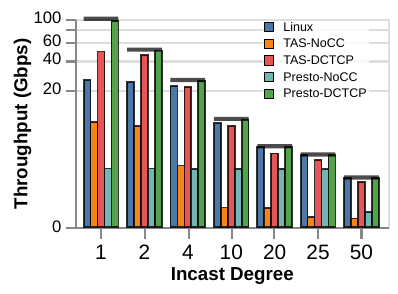}
  \end{subfigure}
  \caption{Streaming goodput: (a) random drops, (b) incast.}\label{fig:stream-drops}\label{fig:incast-response}
\end{figure}

\paragraph{Packet drops}

We measure the streaming throughput while randomly discarding packets with
varying probability. This workload maintains many in-flight packets, ensuring
losses trigger OOO processing. Due to recirculation bottlenecks
(\S\ref{sec:impl}),
we always use a TAS sender. This isolates the evaluation to receiver-side
reassembly. 

Figure~\ref{fig:stream-drops}a shows that \sys{} sustains full throughput up to
a $0.1\%$ drop rate, whereas TAS declines earlier---$2\times$ lower throughput
at $0.1\%$ loss. Both provide 1-OOO reassembly on the receiver,
but \sys{}'s faster acknowledgments accelerate recovery. Linux\footnote{Results
for Linux and the Chelsio TOE can be found in~\cite{flextoe}.} withstands
higher loss rates via selective recovery (e.g., SACK), but cannot reach terabit
throughput even without loss~\cite{cai2021understanding}, while \sys{}
prioritizes fast, bounded-state hardware recovery. The ASIC-based Chelsio TOE
degrades steeply even under moderate loss, and traditional RDMA drops all
out-of-order segments, performing poorly under loss~\cite{mittal2018revisiting,
singhvi2025falcon}.

\paragraph{Congestion control}

To evaluate congestion control, we simulate incast by adding a traffic shaper
frontend to a TAS receiver. The shaper limits ingress bandwidth (corresponding
to the incast degree), uses tail-drop, and marks ECN above a queue occupancy
threshold (determined experimentally as 130). Figure~\ref{fig:incast-response}b
presents the results for a single-connection streaming workload under varying
incast degrees. Black ticks mark the optimal bandwidth achievable at each
degree.

With its control-plane-driven DCTCP and credit-based rate enforcement, \sys{}
effectively regulates transmission, minimizing loss and matching the shaped
line rate. In contrast, \sys{} without congestion control (\sys{}-NoCC)
transmits aggressively and causes excessive tail drops. Even when traffic
shaping is disabled (incast degree=1), the \sys{}-NoCC sender significantly
outpaces the TAS receiver, resulting in up to 50$\times$ lower throughput. TAS
and Linux also regulate their rates effectively but fail to achieve maximum
possible throughput at low incast degrees ($\leq$ 2) due to transport stack
inefficiencies.

\begin{table}[t!]
  \small
  \begin{tabular}{@{}c|rrrr|rrrr@{}}
    \toprule
    \multirow{3}{*}{OOO} & \multicolumn{8}{c}{\textbf{Flow Completion Time (FCT)} [ms]}                                                                                                                                   \\
                         & \multicolumn{4}{c}{Foreground}                               & \multicolumn{4}{c}{Background}                                                                                                  \\
                         & \multicolumn{2}{c}{90p}                                      & \multicolumn{2}{c}{99.9p}      & \multicolumn{2}{c}{90p} & \multicolumn{2}{c}{99.9p}                                            \\
    \midrule
    \multicolumn{9}{@{}l}{\textbf{NewReno}}                                                                                                                                                                               \\
    \midrule
    max                  & 4.97                                                         & ---                            & 17.05                   & ---                       & 68 & ---           & 283 & ---           \\
    0                    & 5.80                                                         & $\times 1.17$                  & 18.49                   & $\times 1.08$             & 91 & $\times 1.34$ & 317 & $\times 1.12$ \\
    1                    & 5.03                                                         & $\times 1.01$                  & 17.05                   & $\times 1.00$             & 84 & $\times 1.23$ & 300 & $\times 1.06$ \\
    2                    & 4.97                                                         & $\times 1.00$                  & 17.05                   & $\times 1.00$             & 78 & $\times 1.15$ & 295 & $\times 1.04$ \\
    \midrule
    \multicolumn{9}{@{}l}{\textbf{DCTCP}}                                                                                                                                                                                 \\
    \midrule
    max                  & 4.69                                                         & ---                            & 12.95                   & ---                       & 62 & ---           & 278 & ---           \\
    0                    & 5.13                                                         & $\times 1.09$                  & 28.52                   & $\times 2.20$             & 71 & $\times 1.14$ & 283 & $\times 1.02$ \\
    1                    & 4.70                                                         & $\times 1.00$                  & 12.91                   & $\times 1.00$             & 62 & $\times 1.00$ & 279 & $\times 1.00$ \\
    2                    & 4.70                                                         & $\times 1.00$                  & 12.98                   & $\times 1.00$             & 63 & $\times 1.01$ & 275 & $\times 0.99$ \\
    \bottomrule
  \end{tabular}
  \caption{Simulated FCTs across OOO tracking fidelity levels.}\label{tab:ooo-simulation}
\end{table}

\paragraph{Large-scale simulation}
We next study large-scale behavior via simulation.
Using TCT's ns-3~\cite{tct} (TCT disabled), we test whether bounded OOO
reassembly and selective recovery suffice for real datacenter loss patterns.
We model a 96-node leaf-spine datacenter topology with latency-sensitive
partition-aggregate incasts of 8KB queries competing with bandwidth-heavy
background flows, drawn from real-world traces~\cite{tct}. We extend ns-3's TCP
to support varying reassembly fidelity, tracking up to $N$ OOO intervals. OOO-0
drops all OOO segments, while OOO-max tracks all segments with unlimited state.

Table~\ref{tab:ooo-simulation} compares the impact of different OOO fidelities.
As expected, the OOO-0/go-back-N baseline performs poorly, greatly inflating
tail FCTs. In contrast, \sys{}'s OOO-1 design with SACK substantially improves
tail FCT, especially under DCTCP, which reacts proactively to congestion.
Beyond 1--2 intervals, fidelity yields diminishing returns: congestion losses
are typically bursty, so a few intervals suffice for efficient SACK-based
recovery. Takeaway: selective recovery is compatible with \sys{}'s RMT
pipeline, and a small, bounded OOO state (1--2 intervals) captures most of the
benefit while remaining implementable on today's hardware.

\paragraph{Performance isolation}
Run-to-completion TCP stacks struggle to isolate latency-sensitive (LS) and
bandwidth-intensive (BI) connections. Specifically, TAS hashes connections to
cores, mixing LS and BI connections, which causes interference, inflating LS
tail latency and reducing BI bandwidth~\cite{cai2021understanding}.

To evaluate performance isolation, we co-locate an echo server RPC workload
with 32 latency-sensitive (LS) connections (single in-flight request) and a
streaming workload with 4 bandwidth-intensive (BI) connections (5 Gbps each).
We first measure both in isolation, then compare with concurrent execution. We
provision TAS with two fast-path cores to avoid bottlenecks.

\begin{figure}[t!]
  \centering
  \includegraphics[width=\columnwidth]{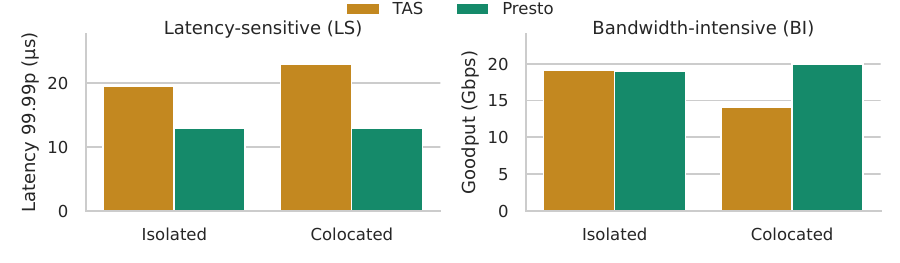}
  \caption{Performance isolation for connections.}\label{fig:perf-isolation}
\end{figure}

As Figure~\ref{fig:perf-isolation} shows, \sys{} delivers optimal performance
for both LS and BI flows: consistent per-packet overhead and independent DMA
channels steer payload directly to application cores, preventing cache
pollution. Under co-location, TAS shows an 18\% higher LS tail latency and 30\%
lower BI bandwidth.

\subsection{Generalizability}\label{sec:eval-fpga}
Prior FPGA work implements TCP as a monolithic circuit, demanding ever-shorter
timing and more power to scale. \sys{} eases this with fine-grained per-state pipelining, even within a single
connection. To investigate whether \sys{} generalizes
to FPGA-based SmartNICs, we port \sys{}'s stateful egress pipeline to the AMD
Alveo U250~\cite{amdalveou200} using the VitisNet P4
compiler~\cite{vitisnetp4}. Our design meets 10ns timing for $300$ Mpps while
using only $2.5\%$ of LUTs and $0.78\%$ of BRAM to support $32$K connections,
consuming $4.93$W worst-case power. The low resource footprint leaves headroom
to replicate the pipeline and shard connections for higher rates and
connections.

By comparison, on a similar FPGA with identical timing constraint,
Tonic~\cite{tonic} peaks at $100$ Mpps and consumes over $20\%$ of LUTs and
BRAM, supporting only $1$K connections. Beehive~\cite{beehive} peaks at $2.6$
Mpps with comparable resource use. Direct comparison is limited since each
design implements a different subset of TCP features---Tonic drops TCP's
byte-stream abstraction in favor of a fixed segment size, while Beehive omits
OOO handling and congestion control but implements connection handshake.
 \section{Related Work}\label{sec:related}

\paragraph{Software stack improvements}
Software TCP stacks, including Linux and kernel-bypass
designs~\cite{tas,arrakis,ix,mtcp,demikernel}, scale by distributing
connections across threads with sequential per-connection
processing~\cite{cai2021understanding,tas}. Despite optimized
interfaces~\cite{affinity_accept,fastsocket,mtcp,megapipe,io_uring,io_uring_man,linuxzerocopy,stackmap},
drivers~\cite{packetmill,niq,netmap,reframer,258955,ccnic}, hardware
assistance~\cite{aRFS,jumbo,about_nics,tso,ddio,ioat}, and
disaggregation~\cite{flextoe,netchannel}, these remain CPU-limited, failing to
exploit network processing's inherent packet-level parallelism. \ \sys{}
instead pipelines TCP state updates across RMT stages even for a single
connection, achieving line-rate processing with ASIC-class efficiency.

\paragraph{Co-designed transports}
Tightly coupled application-transport
designs~\cite{erpc,kalia2016fasst,dragojevic2014farm,homa,r2p2,i10,1rma},
including RDMA~\cite{rdma}, achieve high performance but burden developers with
transport complexities, buffer management, and restrictive execution models,
sacrificing generality and software reuse.\ \sys{} maintains broad TCP and
POSIX compatibility, with flexibility for diverse application needs.

\paragraph{FPGA stacks}
FPGAs enable flexible offload~\cite{accelnet,catapult},
including for TCP\@. Tonic~\cite{tonic} provides a sender-only TCP stack
(evaluated in simulation), while other efforts~\cite{beehive,limago,pigasus} do
not integrate with host applications~\cite{beehive,limago,pigasus}. Partial
offloads like ZeroNIC~\cite{zeronic} and Enso~\cite{enso} accelerate payload
copies but leave TCP on the CPU\@.

These designs implement TCP state processing as monolithic \emph{circuits} with
tight per-packet timing (e.g., 10\ ns in Tonic~\cite{tonic}), making them hard
to extend and power-intensive at high frequencies.\ \sys{} maps TCP to a
pipelined match-action architecture that achieves (\shortsecref{sec:eval-fpga})
higher throughput and lower resource usage than single-circuit designs, while
retaining P4 agility over Verilog~\cite{catapult,kernelinterposeflexible}.

\paragraph{SmartNIC stacks}
SmartNICs are widely used for network
virtualization~\cite{amazon_nitro,google_ipu,intel_ipu,catapult}.
AccelTCP~\cite{acceltcp} and IO-TCP~\cite{io-tcp} partially offload TCP\@;
FlexTOE~\cite{flextoe} fully offloads the TCP data-path, but faces single-core
NPU limits~\cite{chen2024demystifyingdpa,flextoe};
Falcon~\cite{singhvi2025falcon} and SCR~\cite{zhao2025white} pair fixed
transports with NPU congestion control.
Data-parallel decompositions across NPU cores~\cite{snap,netchannel,flextoe}
are bounded by inter-core queueing, per-core variability, and serialized
per-connection logic (FlexTOE's \emph{Protocol} module bottlenecks at $\approx
8$\,Mpps/core~\cite{flextoe}). This serialization is fundamental: even on a
CPU, ZeroNIC~\cite{zeronic} reports a $560$\,Gbps for single-core ceiling at 9K
MTU\@. By pipelining TCP state updates across RMT, \sys{} extracts
instruction-level parallelism, sustaining terabits within a single connection.

\paragraph{RMT transport acceleration}
RMT is emerging as a platform for flexible packet
processing~\cite{pegasus,mind,concordia,redplane}.
NanoTransport~\cite{nanotransport} uses P4 but offloads reassembly and
packetization to FPGAs;\ \sys{} realizes both fully within RMT\@.
R2P2~\cite{r2p2} uses RMT for UDP-based RPC load balancing. GEM implements
lightweight RDMA for external memory access~\cite{gem,tea,ribosome}.
Conweave~\cite{conweave} applies RMT to buffer out-of-order RDMA packets for
network load balancing, and LinkGuardian~\cite{linkguardian} adds link-local
retransmissions. Other proposals leverage RMT for scheduling and congestion
control~\cite{rmtcalendarq,rmtalloc,rmtfairq,bfc}. None provide full transport
offload or obviate the need for an end-to-end reliable protocol like TCP, as
\sys{} does.

\section{Discussion}\label{sec:discussion}
We now turn to \sys{}'s limitations and its
generalizability~\cite{shashidhara2025building}.

\label{sec:limitations}

\paragraph{Datacenter (DC) focus}
\sys{} targets DC environments---Bpps rates, Tbps bandwidth, $\mu$s
latency, ASIC-efficiency, and software flexibility---exploiting their
defining traits: persistent connections, congestion-driven (not
corruption-driven) loss, short RPCs, and benign traffic patterns.
Wide-area networks (WANs) and High-performance computing (HPC)
settings may warrant different trade-offs, yet \sys{}'s principles still
apply: BBR~\cite{bbr} is implementable, and although WAN loss rates may
demand more OOO intervals, \sys{}'s OOO design may track in-flight segments
efficiently for large bandwidth-delay-product (BDP) flows~\cite{omnidma}.

\paragraph{Advanced DC techniques}
Several techniques remain future work. Multi-pathing for AI cluster
congestion~\cite{singhvi2025falcon,zhao2025white} maps onto \sys{}'s OOO
intervals, which tracks cross-path reordering, though we do not evaluate it.
RACK~\cite{ietf8985RACKTLP}, which tracks per-packet transmission times and
scans $O(\text{window})$ unacknowledged packets, is harder given RMT's strict
state limits; but such settings run few large flows per host at modest packet
rates, permitting slower RMT pipelines with stronger per-stage processing.
Network stack disaggregation, enabled by \sys{}'s switch-based prototype, is
also future work.

\paragraph{Message-boundary parsing}
Flexible offloads for higher-level protocols (e.g., RPC load balancing or
access control) that inspect payloads need application-level message
boundaries, which TCP's byte-stream semantics obscure when delimiters straddle
segments (\shortsecref{sec:flex-case-studies}). Pushing message orientation
into the transport solves this.

\paragraph{Head-of-Line (HoL) blocking with \libsyszc{}}
\sys{}'s zero-copy API transmits segments in allocation order,
occasionally causing HoL delays. This rarely matters in practice:
small RPCs can use copy-based APIs cheaply, and
bulk flows can spread across connections.
\sys{} can also expose queue-based interfaces (e.g., \texttt{ibverbs}-style),
via table-lookup address translation.

\paragraph{Generalizability beyond TCP}
\sys{}'s principles extend beyond TCP to many
transports~\cite{shashidhara2025building}. As Tonic~\cite{tonic} observes,
reliable transports share common structural patterns:
tracking in-flight segments for reassembly, recovering losses via
retransmissions, and regulating in-flight segments to avoid congestion,
differing only in mechanism. \sys{} realizes these efficiently on RMT
architecture by overcoming the dependencies of
Figure~\ref{fig:tcp-dependencies}.

Message-oriented transports (e.g.,~RDMA) are a case in point, replacing
byte-stream reassembly with explicit message boundaries---no fundamental
changes to \sys{}'s design. Messages are tracked by packet sequence number
(PSN) rather than byte offsets, and OOO intervals map directly to PSN ranges.
By decoupling data placement from application notification, \sys{} eliminates
on-NIC buffering: each message either carries a host buffer address or draws
from a queue of pre-allocated buffers (cf.~RDMA WRITE vs. SEND), which \sys{}
indexes to place directly into host memory, advancing the base index as the
window moves. We leave implementation and evaluation to future work.
 \section{Conclusion}
\sys{} redefines the design-space of high-performance TCP stacks by showing that
full transport functionality fits within RMT constraints. Embracing pipeline-parallelism,
optimistic concurrency, and pseudo-segment updates, \sys{} delivers terabit
throughput, low latency, and hardware efficiency while preserving software flexibility.
Our Intel Tofino~2 prototype proves that RMT data planes
can support stateful, loss-resilient, and scalable transport, laying a foundation
for next-generation SmartNICs and datacenter networks.
 \begin{acks}
  We thank the anonymous reviewers and our shepherd for their helpful feedback.
  This work is supported in part by NSF grant 2212193 and the University of
  Washington Center for the Future of Cloud Infrastructure (FOCI).
  This work raises no ethical concerns.
\end{acks}
 
\bibliographystyle{ACM-Reference-Format}
\bibliography{paper}

\clearpage
\appendix 
\end{document}